\documentclass[12pt]{article}
\usepackage[top=1in,bottom=1in,right=1in,left=1in]{geometry} 
\usepackage{amsmath,amssymb}  
\usepackage{bm}  
\usepackage{graphicx} 

\graphicspath{ {./Fig/} }
\usepackage{caption}
\usepackage{subcaption}
 \usepackage{color,soul}
 \usepackage{indentfirst}
\usepackage[authoryear]{natbib}
\usepackage{mathtools}
\usepackage[normalem]{ulem}
\usepackage{hyperref}
\hypersetup{
  colorlinks=true,
  citecolor=blue,      
  }
\usepackage{mathptmx, courier, textcomp}
\usepackage[title]{appendix}
\usepackage{soul}
\usepackage{subcaption}
\usepackage{setspace}
\usepackage{placeins}
\usepackage{authblk}
\title{New Pilot-Study Design in Functional Data Analysis}

\author[1]{Ping-Han Huang}
\author[1]{Ming-Hung Kao}
\affil{School of Mathematical and Statistical Sciences, Arizona State University, Tempe, AZ, 85287-1804, USA.}
\date{} 
\affil[*]{Corresponding author. Email address: Ping-Han.Huang@asu.edu}

\begin{document}

\maketitle

\begin{abstract}
Efficient data collection is essential in applied studies where frequent measurements are costly, time-consuming, or burdensome. This challenge is especially pronounced in functional data settings, where each subject is observed at only a few time points due to practical constraints. Most existing design approaches focus on selecting optimal time points for individual subjects, typically relying on model parameters estimated from a pilot study. However, the design of the pilot study itself has received limited attention. We propose a framework for constructing pilot-study designs that support both accurate trajectory recovery and effective planning of future designs. A search algorithm is developed to generate such high-quality pilot-study designs. Simulation studies and a real data application demonstrate that our approach outperforms commonly used alternatives, highlighting its value in resource-limited settings. \\

\end{abstract}

\maketitle

\section{Introduction}
One classical assumption of functional data analysis (FDA) is that the observations are collected densely in time \citep{ramsay1982data,silverman1985some,rice1991estimating}. However, in practice, there are often constraints limiting the collection of dense data. One key limitation is the measurement cost and burden associated with the frequent data collection. For instance, in medical studies, obtaining frequent blood samples from patients can be impeded by high costs and patient burden \citep{ lopes2021real,pan2023reinforced}. Additionally, in fields like environmental sciences, measuring devices such as satellite imaging might be available only once every few days due to technical limitations \citep{zhu2022spatiotemporal, gregory2024scalable}. Subject availability also contributes to this issue, as participants in clinical studies may be unable to adhere to frequent data collection schedules. This often leads to missing values, introducing additional complexity to the analysis \citep{shi2021functional, kodikara2022statistical}. 

Building on this recognition, previous studies in optimal designs for sparse functional data focused on finding the optimal time points to collect measurements from subjects to be enrolled in the next study. In order to find the optimal next-study design, prior information is needed and can be obtained from a pilot data set. For instance, \citet{ji2017optimal} obtained estimates of unknown parameters from a pilot data set and used these estimates in two design optimality criteria that they considered: one for minimising the mean integrated squared error for recovering the underlying random function, and the other for minimising the prediction error of scalar responses in functional linear regression. Following \citeauthor{ji2017optimal}'s work, \citet{park2018joint} targeted the same goals and further developed a unified joint optimality criterion for selecting optimal time points that strike a balance between the two objectives. They compared the functional models to classical mixed effects models and demonstrated that functional models are preferred for prediction-based designs. 

In the same vein, \citet{rha2020design} developed flexible weighted sum criteria by considering the design efficiency for trajectory recovery and that for response prediction with function-on-function regression models. They further built a probabilistic subset search (PSS) algorithm to find optimal designs. Utilising the estimated between-subject variability from a pilot data set, the PSS algorithm is demonstrated to be more efficient than its competitors in the search of optimal designs, by allocating higher selection probability to time points with higher between-subject variability.

The aforementioned optimal design approaches heavily rely on the amount of information about the relevant model parameters, provided by the pilot data set. The importance of having a well-designed pilot study became evident through a motivating example that first drew our attention to this issue. \citet{zhong2022robust} conducted a longitudinal study of CD4 immune cells based on a data set of 190 subjects collected from 1997 to 2002. The data set showed a significant dearth of observations collected across subjects in the first 400 days of the study. It caused a void of information that is required for their data analysis. Consequently, they discarded all the data points from the first 400 days in data pre-processing step. The lack of design consideration resulted in data that were inadequate for accurate estimation, highlighting a broader and often overlooked problem in sparse functional data analysis. 

A good pilot-study design, which gives the best set of time points for sampling each random curve in the pilot study, would have avoided the previously mentioned issue in the first place. In this direction, our study focuses on developing a good pilot-study design to render a precise pilot-study inference to allow the investigators to successfully  identify the ensuing optimal next-study designs for future data collection. In particular, the statistical inference that we consider here is on recovering the trajectory of the underlying random function, which is especially important and challenging for sparse functional data. With this consideration, our target is at a good pilot-study design to facilitate the use of the previously mentioned design approaches in the  search of an optimal next-study design. In addition, we would like our selected pilot-study design to have a reasonably good statistical efficiency in  recovering the random curves in the pilot study. To our knowledge, research on selecting such pilot-study designs is currently unavailable.

Here, we propose a new design structure and a search algorithm for finding a good pilot-study design. Our main idea is to construct the design by considering a hybrid structure that brings in the strengths of snippet \citep{galbraith2017accelerated} and balanced incomplete block designs. A discussion of related design concepts can be found in Section \ref{2.2} and references therein. Our algorithm for constructing such a pilot-study design is based on linear integer programming. By applying to real data application, we show that our new design facilitates generating high-quality FDA designs for subjects in the next study and has a good statistical efficiency in trajectory recovery in the pilot study.

The rest of paper is organised as follows. In Section \ref{sec model and design}, we introduce FDA methods that help recover the underlying trajectory for sparse functional data. We also review relevant existing design structures along with their advantages and limitations. In Section \ref{sec hybrid design}, we propose a new pilot-study design and introduce an algorithm to generate such a design. Further, we develop an optimality criterion for evaluating the pilot-study design performance. In Section \ref{sec real data}, we apply our proposed approach to a real-world case study on age-related patterns of fecundity for female Mediterranean fruit flies, where we describe the study context, discuss the motivation of the design problem, and demonstrate its practical effectiveness. Finally, Section \ref{sec:discussion} concludes with a summary of key findings. 

\section{FDA Model and Related Designs} \label{sec model and design}

\subsection{Underlying Model} \label{2.1}
In the application of sparse FDA, the number of repeated measurements is limited due to practical constraints such as budgets and financial costs. The measurements are usually contaminated with errors and collected at irregularly spaced time points. In modelling such data, \citet{yao2005functional} proposed the following model:
\begin{equation} \label{eq:model}
  U_i(t_{ij}) = X_i(t_{ij})+e_{ij} \text{, } i=1,\ldots, n \text{ and } j=1, \ldots, K,
\end{equation}
where $X_i$ is the $i^\text{th}$ random curve defined over a closed and bounded time interval $\mathcal{T}$, $U_i(t_{ij})$ is the $j^\text{th}$ noisy observation of $X_i$ at time $t_{ij} \in \mathcal{T}$, and $e_{ij}$ is the measurement error that follows the normal distribution with zero mean and variance $\sigma_e^2$. It is assumed that $X_i$'s are i.i.d. square-integrable random functions that have a continuous mean function $\mu(t)$ and continuous covariance function $\bm{\Sigma}$. Further, the errors $e_{ij}$ are i.i.d. and are independent of  $X_i$. 

To recover the underlying trajectory  of each $X_i$, functional principal component analysis (FPCA) has been a popular tool. Specifically, by Mercer's theorem, there exists a set of orthonormal eigenfunctions $\psi_m(t)$'s and the corresponding non-increasing eigenvalues $\lambda_1 \ge \lambda_2 \ge \ldots \ge 0$ such that the covariance function can be written as $\sum_{m=1}^\infty \lambda_m \psi_m(t) \psi_m(s)$. Moreover, with the Karhunen-L\'{o}eve representation, we may re-write model \eqref{eq:model} as \eqref{eq:FPCA}.
    \begin{equation} \label{eq:FPCA}
        U_i(t_{ij}) = X_i(t_{ij})+e_{ij} = \mu(t_{ij}) + \sum_{m=1}^{\infty} \xi_{im} \psi_{m}(t_{ij}) + e_{ij},
    \end{equation} 
where $\xi_m = \int \{X(t) - \mu(t)\} \psi_m(t) \,dt$ are the uncorrelated functional principal component (FPC) scores with mean 0 and variance $\lambda_m$.

With sparse functional data, the traditional numerical integration approach for estimating FPC scores no longer works. In view of this issue, \citet{yao2005functional} proposed the Principal Components Analysis through Conditional Expectation (PACE) method that uses pooled functional data across subjects and local linear smoothers to estimate the model. Under the Gaussian assumptions, we may obtain the estimates of FPC scores $\hat{\xi}_{im}$ as

\begin{equation} \label{eq:PACE}
   \hat{\xi}_{im} = \hat{E}[\xi_{im} | \bm{U}_{i}] = \hat{\lambda}_{m} \hat{\psi}_{im}^T \hat{\bm{\Sigma}}_{\bm{U}_{i}}^{-1} (\bm{U}_{i} - \hat{\bm{\mu}}_i),
\end{equation}
where $\bm{U}_{i} = (U_i(t_{i1}),\ldots, U_i(t_{iK}))^T$, $\bm{\mu}_i = (\mu(t_{i1}),\ldots,\mu(t_{iK}))^T$, $\psi_{im} = (\psi_m(t_{i1}),\ldots, \psi_m(t_{iK}))^T$, $\bm{\Sigma}_{\bm{U}_{i}} = Cov(\bm{U}_{i},\bm{U}_{i}) = Cov(\bm{X}_{i},\bm{X}_{i})+\sigma_e^2\bm{I}_{K}$, and $\bm{X}_{i} = (X_{i}(t_{i1}),\ldots, X_{i}(t_{iK}))^T$. The estimates of the unknown quantities in \eqref{eq:PACE} are obtained with the smoothing methods described in \cite{yao2005functional}. With \eqref{eq:PACE}, the recovery (or prediction) of $X_i(t)$ is often based on the first $M$ leading eigenfunctions as below. 
\begin{equation*}
    \hat{X}_i (t) = \hat{\mu}(t) + \sum_{m=1}^{M} \hat{\xi}_{im} \hat{\psi}_{m}(t).
\end{equation*}

Based on these results, we construct optimality criteria for evaluating the performance of pilot-study designs, as to be explained in detail in Section \ref{sec hybrid design}.

\subsection{Related Designs} \label{2.2}
For ease of exposition, we fix the total number of observations collected from each random curve  to $K$, as in Model \eqref{eq:model}. A regular time grid $T_g= \{\tau_1, \tau_2, ...\tau_v \}$ with $\tau_j < \tau_{j+1}$ is imposed on the compact time domain $\mathcal{T}$ of the random curve. We also consider a realistic situation where only one noisy observation from a curve can possibly be taken at each time point. The total number of subjects in the pilot study is also assumed given at the design stage. With this setup, the search of finding a good pilot-study design reduces to finding the best time points for making observations from each random curve. 

\begin{figure}[h]
\centering
\includegraphics[width=.9\textwidth]{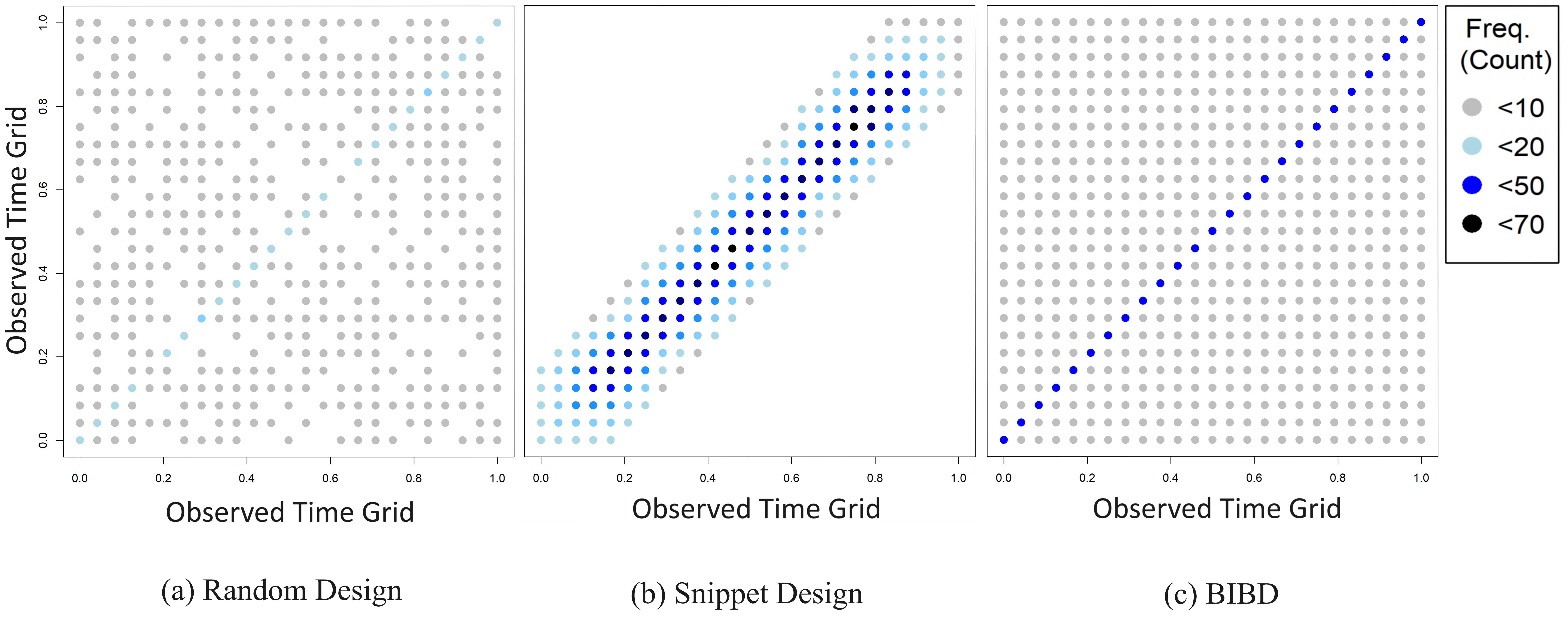}
  \caption{Design plots for (a) a random design, (b) a snippet design, and (c) BIBD.}
  \label{fig:three_plots}
\end{figure}

To visualise a plot-study design, we may draw a design plot with x-axis and y-axis being the time grid $T_g$. The design plot is a scatter plot, on which each point represents an assembled pair of time points ($t_{ij}$,$t_{ik}$) where $t_{ij}$ and $t_{ik}$ are, respectively, the $j^\text{th}$and $k^\text{th}$ time points selected for the $i^\text{th}$ subject. Such a design plot is widely used in FDA; see, e.g., \citet{yao2005functional}. We present the design plots for three different design structures in Figure \ref{fig:three_plots}. There, the colour shade indicates the frequency of assembled time point pairs ($t_{ij}$,$t_{ik}$) across $i$.

The first and perhaps the most straightforward design structure is the random design, consisting of sets of $n$ independent random samples, each of size $K$, from the time domain $T_g$. Figure \ref{fig:three_plots}(a) presents the design plot of a random design. For random designs, the locations of the points, and hence the ``holes'', are random and may differ each time we generate the design. While it is simple to construct a random design, such a design cannot guarantee a good design performance and may further hinder the stability of statistical inferences.

Another possible design structure is the snippet design which is not uncommon in longitudinal studies \citep{galbraith2017accelerated}. A snippet design takes observations at consecutive time points within a short period to minimise the duration of data collection for each subject. As illustrated in Figure \ref{fig:three_plots}(b), points on the design plot are concentrated on the diagonal band and no information is provided outside the diagonal band. While the information provided on the diagonal band helps estimate the error variance, the absence of information creates problems in estimating the covariance function of observations that is however essential for many FDA methods. With snippet designs, the covariance function can easily become non-estimable without regularisation and/or imposing strong assumptions. This poses challenges in modelling functional data \citep{lin2021basis}. 

For constructing a good pilot-study design, we also consider borrowing the concept of balanced incomplete block designs (BIBD) from combinatorial design theory \citep{yates1936incomplete}. To the best of our knowledge, the use of BIBD or related designs is new to FDA. We will detail how to incorporate BIBD in a design for functional data in the next section. But what motivates us to borrow such a concept is that the BIBD follows strict constraints that every pair of treatments occurs together within a block an equal number of times. In the context of FDA, we view the sampling time points as treatments, and the subjects (or the random curves $X_i$'s) as blocks. The previously described feature of BIBDs would then form a uniformly space-filling design plot with no missing information as in Figure \ref{fig:three_plots}(c). Intuitively, this feature will facilitate the estimation of covariance function. However, the existence of a BIBD is not always guaranteed, which could impede the execution of some experiments. Our experience also suggests that the BIBD, when used alone as a pilot-study design for FDA, does not always outperform its competitors (see Section \ref{sec real data}). A further development of methods for constructing designs that possess desired features is thus called for.  In the next section, we propose a new hybrid design structure that tends to strikes the balance between the advantages and drawbacks of the previously mentioned designs to increase the overall design efficiency.

\section{New Design, Search Algorithm, and Optimality Criterion} \label{sec hybrid design}
Built on the previously discussed design concepts, we propose a new FDA design structure for pilot studies to achieve our goals of 1) facilitating the identification of a high-quality FDA design for the next study, and 2) rendering a good prediction of the trajectories of all the underlying $X_i$'s within the pilot study. We then propose a search algorithm for producing such a good pilot-study design and construct an optimality criterion to evaluate the design performance.

Combining the best of two worlds, we would like our pilot-study design to possess the desired features of BIBD and snippet designs. As described previously, the balanced structure of a  BIBD is advantageous for estimating covariance function. By focusing on the diagonal band of the design plot, the snippet design tends to provide additional amount of information for estimating the error variance, $\sigma_e^2$. A shown in \eqref{eq:PACE}, the covariance of the data can be expressed as $Cov(U_{ij}, U_{ik}) = Cov(X_i(t_{ij}), X_i(t_{ik})) + \sigma^2_e\delta_{jk}$, where $\delta_{jk}$ is a Kronecker delta. By having a near-balanced structure in the design plot (as a BIBD) while allocating additional points around the diagonal band with a snippet structure, a hybrid design is expected to provide a good estimation of the covariance of $U_{ij}$'s, which is essential for many inferences in FDA.

To detail the hybrid structure of our new design, we consider an $n \times v$ incidence matrix $\bm{N}$, where $n$ is the number of subjects as in \eqref{eq:model} and $v$ is the size of the time grid $T_g$. Each entry $x_{ij} \in \{0,1\}$ of $\bm{N}$ tells whether an observation is to be drawn from the $i^\text{th}$ subject at the $j^\text{th}$ time point of the time grid. Each row of $\bm{N}$ sums to $K$. Subsequently, the $v \times v$ concurrence matrix $\bm{N}'\bm{N}$ records how many times each pair of time points from $T_g$ are selected together for the same subject. We note that the previously introduced design plot can be viewed as a pictorial representation of the concurrence matrix. The rows of the corresponding incidence matrix of a hybrid design are divided into a snippet portion and a BIBD portion. A constant $w \in (0,1)$ is used to indicate the proportion of rows in $\bm{N}$ that are reserved for the snippet structure. In the first $nw$ rows of the incidence matrix (snippet portion), each subject has a cluster of two consecutive time points plus $(K-2)$ randomly selected time points from $T_g$. Our experience suggests that having a greater number $(> 2)$ of consecutive time points does not seem to improve the design efficiency. The rest $(n - nw)$ rows (BIBD portion) mimics the BIBD structure under relaxed constraints that requires only the concurrence matrix being ``nearly'' completely symmetric. Below, we propose a computer algorithm for constructing such a hybrid design based on the incidence and concurrence matrices introduced. 

Previously, \citet{mandal2014efficient} proposed an algorithm to construct designs by deciding, for each treatment (in our case, time point), a subset of subjects who will receive it. Their algorithm then sequentially runs through all the $v$ treatments, and if successful, it gives a design possessing the desired structures (such as a BIBD). Although the algorithm by \citet{mandal2014efficient} helps to construct BIBD and related designs, applying it to construct our hybrid design does not seem as straightforward. This calls for the development of a new algorithm. In contrast to \citet{mandal2014efficient}'s approach, our algorithm selects $K$ time points for each subject and the procedure is repeated for the $n$ subjects. But with the special design structure that we need, we impose different constraints for different subjects. In other words, our algorithm constructs a hybrid design by sequentially augmenting the individual design for each subject.

For clarity, we may separate our search of a hybrid design into two stages. The first stage is to identify a target concurrence matrix for the given set of parameters $(n,v,K)$. Such a concurrence matrix $\bm{N}'\bm{N}$ resembles a Toeplitz-like matrix structure in \eqref{eq:concurrence of hybrid} that has the same value $c_1$ along the diagonals and the value decreases to $c_2$ and $c_3$ when we move away from the diagonal. 
\begin{equation} \label{eq:concurrence of hybrid}
    \bm{N}'\bm{N} = \begin{pmatrix}
        c_1 & c_2 & c_3 & c_3 &   & c_3\\
        c_2 & c_1 & c_2 & c_3 &  \cdots & c_3\\
        c_3 & c_2 & c_1 & c_2 &   & c_3\\
        c_3 & c_3 & c_2 & c_1 &   & c_3\\
           & \vdots & &   &\ddots&   c_2\\
        c_3 & c_3 & c_3 & c_3 & c_2 & c_1
    \end{pmatrix}.
\end{equation}
Our target values for $c_1$, $c_2$ and $c_3$ are in the following:
\begin{equation} \label{eq:target}
  \begin{split}
    c_1=\frac{n \times K}{v} \text{;} \hspace{0.5em}
    c_2=w \times \overbrace{\frac{n(K-1)}{v-1}}^{\text{snippet}} + (1-w) \times \overbrace{\frac{nK(K-1)}{v(v-1)}}^{\text{BIBD}} \text{; and} \hspace{0.5em}
    c_3=\frac{ n {K\choose2} - c_2(v-1)}{{n-1 \choose 2}}.
  \end{split}
\end{equation}

Here, $c_1$ is computed so that all the time points on the grid appear equally often in the design, and $c_2$ is the target number of appearances of $(\tau_j, \tau_{j+1})$. With a predetermined weight $w$, $c_2$ is calculated by assuming that the first $nw$ subjects are with a snippet structure, and the rest $(n-nw)$ subjects are with a near BIBD structure (even when the corresponding BIBD does not exist). Finally, we would like the remaining off-diagonal entries to have an equal value $c_3$, which is the number of appearances of each time pair $(\tau_j, \tau_k) \in T_g^2$ with $k > j+1$. It is essential to acknowledge that a design whose concurrence matrix is exactly \eqref{eq:concurrence of hybrid} might not exist, but we use \eqref{eq:concurrence of hybrid} as guidance for constructing a good hybrid design. For convenience, we say that the first $nw$ subjects form the ``snippet'' portion of the design. Note that it is required that $nw$ be an integer. In the case where $nw$ is not an integer, we round it to the nearest whole number. We also say that the remaining $(n - nw)$ subjects form the ``BIBD'' portion of the design, although we allow this portion to have some departure from the exact BIBD structure.

In the second stage, we find the optimal hybrid design by building an incidence matrix $\bm{N}$, attempting to make the corresponding values in $\bm{N}'\bm{N}$ to attain the values of our target concurrence matrix computed in the first stage. We start with the first subject having a snippet structure of two consecutive time points, plus $(K-2)$ randomly selected time points. Specifically, the hybrid design is built with an objective function \eqref{eq:obj} and constraints \eqref{eq:k=5} to \eqref{eq:make up}.

\begin{subequations}
  \label{eq:optim}
  \begin{align}
    \text{Max} \hspace{4em}
    &\sum_{j=1}^v w_{ij}x_{ij} \text{ for } i=1, \ldots n\label{eq:obj}\\
    \text{subject to} \hspace{2em}
    & x_{ij} \in \{0,1\} \text{, } \sum_{j=1}^v x_{ij} = K \label{eq:k=5}\\
    & w_{ij} = \min\{\frac{1}{r_{ij}}, 1\} \text{, } 
    r_{ij} = \sum_{\ell=1}^{i-1} x_{\ell j} \label{eq:weight}\\
    & \sum_{i=1}^{n} x_{ij}x_{ij} \le c_1 + \delta \text{ for } j=1, \ldots, v \label{eq:attain c1}\\
    & \sum_{i=1}^{n} x_{ij}x_{i,j+1} \le c_2 + \delta \text{ for } j=1, \ldots, v-1 \label{eq:attain c2}\\
    & \sum_{i=1}^{n} x_{ij}x_{i,j+r} \le c_3 + \delta \text{ for } j=1, \ldots, v-2 \text{ and } r=2, \ldots, v-j  \label{eq:attain c3}\\
     & x_{i,i} + x_{i,i+1} = 2 \text{ for } i=1, \ldots, nw \label{eq:consecutive points}\\
    & \sum_{\ell=1}^{\Delta} x_{i,i-\ell} + x_{i,i+1=\ell} = 0 \text{ for } i=1, \ldots nw \label{eq:snippet jump}\\
     & x_{ij}x_{ij+1} = 0 \text{ for } i=nw+1, \ldots, n \label{eq:no consecutive points}\\
    & \sum_{j=1}^v s_{qj}x_{ij} < K \text{ for } q=1,\ldots, p \label{eq:make up}
  \end{align}
\end{subequations}

The objective function aims to maintain approximately the same values on the diagonal of the resulting concurrence matrix. Even though the location of the two consecutive time points for each subject with snippet structure are fixed, the rest ($K-2$) randomly selected time points per subject remain to be determined, just as the time points for subjects with BIBD structure. To determine the locations for these time points, we follow a rule of thumb that encourages the algorithm to assign an 1 to those time points having a relatively small number of appearances in the previous rows.

All subjects are required to satisfy constraints \eqref{eq:k=5} to \eqref{eq:attain c3}, and \eqref{eq:make up}. Constraint \eqref{eq:k=5} guarantees that each subject has $K$ observations. Constraints \eqref{eq:attain c1}, \eqref{eq:attain c2}, and \eqref{eq:attain c3} are for controlling the values of the resulting concurrence matrix. In particular, $\delta$ there stands for the tolerance level between the target concurrence matrix and the actual values yielded by the algorithm. It is noteworthy that, by selecting a tolerance level $\delta$, the obtained hybrid design is allowed to have a slightly different concurrence matrix than \eqref{eq:concurrence of hybrid}. The choice of $\delta$ would depend on the values of design parameters ($n,v,K$) and is pertinent to algorithm performance. We suggest a small value of $\delta$ so that the resulting concurrence matrix is close to the targeted one and the algorithm remains computationally efficient. However, one may consider increasing $\delta$ when the algorithm fails to identify a design within a given amount of time. 

When running the algorithm, we decide the consecutive time points for the subjects in the snippet portion ($i=1, \ldots, nw$) following constraint \eqref{eq:consecutive points}. 
The other $x_{ij}$'s with $j > 2$ for these $nw$ subjects are then assigned to satisfy constraint \eqref{eq:snippet jump}. In constraint \eqref{eq:snippet jump}, we set $\sum\nolimits_{\ell=1}^{0} x_{\ell j}=0$ and $x_{ij}=0 \text{ for } j<1 \text{ or } j>v$. This second constraint for the snippet portion of the design prevents a large cluster of time points within the same subject by keeping the remaining $x_{ij}$'s sufficiently distant (with at least $\Delta$ grid points away) from the cluster of consecutive time points. The choice of $\Delta$ depends on the time grid size $v$ and the number of observations per subject $K$. We suggest a small value of $\Delta$ so as to reduce the chance of having another cluster of time points. For the BIBD portion, our algorithm searches for $x_{ij}$ values with constraint \eqref{eq:no consecutive points}. This constraint helps to avoid destroying the design structure that have already been formed  in the snippet portion. That is, no subject in the BIBD portion is allowed to take observations from consecutive time points.

In the case where a feasible solution cannot be found for a specific subject (a specific row in $\bm{N}$), we adapt a similar approach of \citet{mandal2014efficient}. Specifically, we randomly select a subject without replacement from the previous ones, and regenerate a solution (i.e., an individual design) for the selected subject. If a solution of the selected subject is found, the make-up process ends and the algorithm moves on to find solutions for rest of the subjects. Otherwise, this make-up process is repeated until a predefined maximal number of iterations is reached. To avoid a repeated use of the same solutions, a matrix $\bm{S}$ is constructed (with default null values) to store the original solutions of the subjects that are selected for solution regeneration. Constraint \eqref{eq:make up} is then imposed to prevent future subjects from getting a solution that is already in $\bm{S}$. In constraint \eqref{eq:make up}, $s_{qj}$ represents $(q, j)^{\text{th}}$ entry of $\bm{S}$ and $p$ is the number of rows in $\bm{S}$. We refresh $\bm{S}$ by setting it to empty when $p>2N$. 

We evaluate our pilot-study designs in achieving two study goals. The primary goal is to facilitate the search of high-quality FDA designs for the next study. The secondary objective is to increase the statistical efficiency in recovering trajectories of $X_i$'s for the subjects in the pilot study. To evaluate the performance of pilot-study designs, we develop a composite criterion consisting of two objective functions corresponding to these two goals. We begin by considering the mean integrated squared error (MISE) of the best predictor for an $X_{i'}$ in the next study. The MISE is presented below as a function of the individual design $\bm{t}=(t_1, \ldots, t_K) \subset T_g$. 
\begin{align} \label{eq:prediction error}
  MISE(\bm{t}) &= E \int_{\mathcal{T}} [X_{i'}(t) - E[X_{i'}(t)|U_{i'}(\bm{t})]]^2 \,dt \nonumber \\
  &= tr\{\hat{\bm{\Lambda}}\} - tr\{\hat{\bm{\Lambda}}\hat{\bm{\Psi}}(\bm{t})' [\hat{\bm{\Psi}}(\bm{t})\hat{\bm{\Lambda}}\hat{\bm{\Psi}}(\bm{t})'+\hat{\sigma}_e^2\bm{I}_k]^{-1} \hat{\bm{\Psi}}(\bm{t})\hat{\bm{\Lambda}}\}.
\end{align}
where $U_{i'}(\bm{t}) = (U_{i'}(t_{1}), \ldots, U_{i'}(t_K))'$, $\hat{\bm{\Lambda}} = diag(\hat{\lambda}_1, \ldots, \hat{\lambda}_m)$, and $\hat{\bm{\Psi}}(\bm{t}) = (\hat{\psi}_1(\bm{t}), \ldots, \hat{\psi}_m(\bm{t}))$. As suggested in the previous studies, the parameter estimates involved in \eqref{eq:prediction error} are obtained from the pilot study. With this, and the fact that the first term in \eqref{eq:prediction error} does not depend on $\bm{t}$, we rewrite the criterion for evaluating the next-study design as an estimated $\hat{F}(\cdot)$.
\begin{equation} \label{eq:Fhat}
  \hat{F}(\bm{t}) = tr\{\hat{\bm{\Lambda}}\hat{\bm{\Psi}}(\bm{t})' [\hat{\bm{\Psi}}(\bm{t})\hat{\bm{\Lambda}}\hat{\bm{\Psi}}(\bm{t})'+\hat{\sigma}_e^2\bm{I}_k]^{-1} \hat{\bm{\Psi}}(\bm{t})\hat{\bm{\Lambda}}\}.
\end{equation}
This $\hat{F}(\bm{t})$ has been used in, e.g., \cite{park2018joint} and \cite{rha2020design}, to obtain an optimal $\bm{t}_{opt}$ for a new subject in the next study. Without a good pilot-study design, the estimated $\hat{F}(\bm{t})$ can be poor, leading to an inefficient $\bm{t}_{opt}$. To compare the performance of pilot-study designs in obtaining a good $t_{opt}$, we calculate the following absolute relative error (ARE).
\begin{equation} \label{eq:ARE}
    \text{ARE} = (\lvert F(\bm{t}_{opt}) - F(\bm{t}^*) \rvert) / F(\bm{t}^*).
\end{equation}
Again, the optimal $\bm{t}_{opt}$ is obtained by maximising the estimated $\hat{F}(\cdot)$. The true optimal $\bm{t}^*$ is by maximising the true $F(\cdot)$. Both $\bm{t}_{opt}$ and $\bm{t}^*$ are obtained by a pre-specified search algorithm such as the PSS algorithm proposed by \cite{rha2020design} or other search algorithms.

For our second objective, the performance of a pilot-study design will be evaluated by the following relative root mean squared error (RRMSE) for the subjects in the pilot study. 
\begin{equation} \label{eq:RRMSE}
  \text{RRMSE} = \sqrt{\frac{1}{n}\sum_{i=1}^{n} \sum_{j=1}^{K} \{X_i(t_{ij}) - \hat{X}_i(t_{ij})\}^2 \bigg/ \sum_{j=1}^{K} X_i(t_{ij})^2}. \nonumber
\end{equation}
To strike a balance between the two goals, a weighted sum is taken to combine the two criteria. We note that the weights for the two criteria can be altered if one of the goals is preferred to the other. Our composite criterion is thus:
\begin{equation} \label{eq:composite criterion}
  \begin{split}
    \text{Criterion} = 0.5 \times \text{ARE} + 0.5 \times \text{RRMSE}.
  \end{split}
\end{equation}
In the next section, we will utilise this composite criterion to compare the performance of our hybrid design to other designs.

\section{Simulation and Results} \label{sec simulation}
In this simulation study, we aim to evaluate and compare our hybrid design performance to BIBDs and random designs and demonstrate that our hybrid design serves as a good pilot-study design that facilitates the search of high-quality FDA designs for the next study, while providing sufficient information for the estimation of the pilot data set. The corresponding R codes are available upon request to the first author.

\subsection{Simulation Settings} \label{sec simulation setting}

We generate 10 functional data sets by adopting the model framework \eqref{eq:FPCA} from \citet{yao2005functional}. Without loss of generality, we assume $t \in \mathcal{T} = [0,1]$. Following the previous works \citep{yao2005functional, park2018joint, rha2020design}, we discretise $\mathcal{T}$ into 25 equally spaced time points. Each underlying random function of interest is constructed by $X_i(t) = \mu(t) + \sum_{m=1}^5 \xi_{im}\psi_{m}(t)$, where $\mu(t)= t + \sin{(t)}$ and $\xi_{im}$'s are independently generated from normal distributions $\mathcal{N}(0,\lambda_{m})$ with $\lambda_1 \ge \lambda_2 \ge \ldots \ge \lambda_5$. We $\psi_1 (t) = \sqrt{2} sin(2\pi t) \text{, } \psi_{2}(t) = \sqrt{2} cos(2\pi t) \text{, } \psi_{3}(t) = \sqrt{2} sin(4\pi t) \text{, } \psi_{4}(t) = \sqrt{2} cos(4\pi t)  \text{ and } \psi_{5}(t) = \sqrt{2} sin(6\pi t)$ and the corresponding eigenvalue $\lambda_{m}$ equals $10/2^m$ with $m=1,\ldots,5$. The noisy observations are obtained by $U_i(t_{ij}) = X_i(t_{ij})+e_{ij}$, where $e_{ij}$ is generated from the normal distribution $\mathcal{N}(0,\sigma_e^2)$. Here, the error variance $\sigma_e^2$ is set to $0.9688$ such that the signal-to-noise ratio $\sigma_e^{-2} \sum_{m=1}^5 \lambda_m$ equals $10$. 

After obtaining the 10 functional data sets, we ``sparsify'' each data set based on three different design structures, including BIBD, random, and our hybrid design, to get sparse pilot data sets with $K=5$ observations per subject. Furthermore, to compare the designs with different scenarios, the above steps are repeated for different numbers of subjects, namely $n=50,60,\ldots,240$. Finally, we apply the PACE method to obtain estimates of subject trajectory and use the composite criterion \eqref{eq:composite criterion} proposed in Section \ref{sec hybrid design} to evaluate the performance of each design. 

In constructing the sparse pilot data sets, there are two issues we would like to note here. First, as each design structure involves some degree of randomness in the design generating process, we generate 20 designs for each design structure in order to study the stability of design performance. In particular, for the BIBD structure, we generate 20 isomorphic BIBDs. This is done by randomly relabelling the treatments of a design. Note that although the two isomorphic designs are ``equivalent'' in most classical design settings, they can have different design efficiencies in the FDA setting. Therefore, for each functional data set and each design structure, we have 20 sparse pilot data sets. The second issue is that, for computation simplicity, the BIBD parameters are set to $(n,v,K,r,\lambda) = (30,25,5,6,1)$. 
To construct BIBDs with different numbers of subjects, we replicate the existing BIBD in multiples of 30, plus randomly selected subjects from the existing BIBD to fulfil the required number of subjects.

\subsection{Design Performance on Reaching Two Goals} \label{sec simulation first}
With the above simulation settings, we compare the three designs using the composite criterion \eqref{eq:composite criterion} in Section \ref{sec hybrid design}. Our simulation results are summarised in Figure \ref{fig:PSS result}. For each number of subjects that we consider, we present in Figure \ref{fig:PSS result} the box plots of the criterion values for the three design structures. Each box summarises the averaged criterion values over the 10 functional data sets for the 20 designs for each design structure.

\begin{figure}[h]
\centering
\includegraphics[width=.90\textwidth]{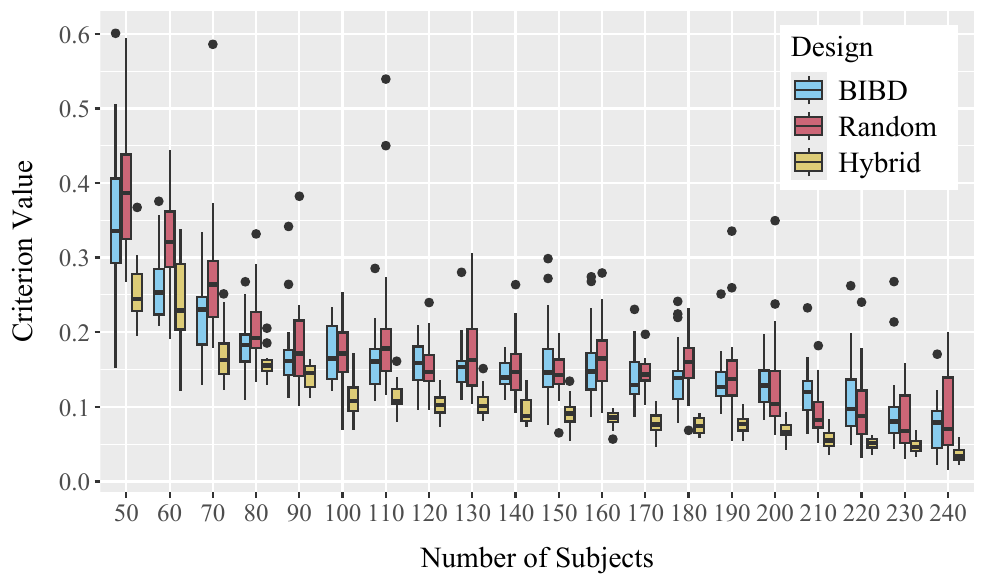}
  \caption{Grouped box plot of designs for composite criterion.}
  \label{fig:PSS result}
\end{figure}

For all the scenarios we studied in Figure \ref{fig:PSS result}, the performances of our hybrid design consistently surpasses the other two designs in terms of both the median and interquartile range. While the BIBD and random design have similar median values, random design normally has a larger interquartile range. It is noteworthy that the median and interquartile ranges of all designs tend to increase as the number of subjects decreases. 

\subsection{Design Robustness with General Search Algorithm} \label{sec simulation second}
In the previous section, we applied the PSS algorithm to find the optimal $\bm{t}_{opt}$ and $\bm{t}^*$ by maximising $\hat{F}$ and $F$, respectively; see also Section \ref{sec hybrid design}. Some other search algorithms, such as exhaustive search or sequential selection algorithm \citep{ji2017optimal}, may also be considered. These algorithms are prone to offer different efficiencies, which could potentially impact the performance of different pilot-study designs. An additional simulation is therefore conducted here to further verify the usefulness of our hybrid design, specifically in determining designs in the next study with different search algorithms. In this section, we assume arbitrary search algorithms with different abilities in obtaining optimal designs for subjects in the next study. For example, some algorithms might be able to find the optimal solution maximising $\hat{F}$, while others can only attain a solution having, say, 95\% efficiency of the optimal solution (under the same $\hat{F}$).

We adopt the same settings as in Section \ref{sec simulation setting} to generate 10 functional data sets, 20 pilot-study designs for each of the three design structures (BIBD, random, and hybrid designs) and the corresponding sparse pilot data sets. Then arbitrary search algorithms for finding the next-study design are assumed to at least achieve certain efficiency threshold $\theta \in \{99\%, 97\%, 95\%\}$, relative to the optimal solution. We note again that the true objective function $F(\cdot)$ is not available in practice, and these search algorithms can only work with the surrogate $\hat{F}(\cdot)$ in \eqref{eq:Fhat}, with the involved unknown quantities estimated from the pilot study. A design that achieves $100\%$ efficiency under $\hat{F}(\cdot)$ does not necessarily achieve $100\%$ efficiency under $F(\cdot)$. In light of this, for each threshold (i.e. efficiency level) $\theta$, we evaluate the efficiency $\text{eff}_{\hat{F}} (\bm{t})= \hat{F}(\bm{t})/ \hat{F}(\bm{t}_{opt})$ for all of the $v \choose K$ $=$ 531,130 possible candidate designs and then select all designs having $\text{eff}_{\hat{F}}(\bm{t}) \geq \theta$

The above procedure will give three groups of designs $\bm{t}$'s, each for an efficiency level $\theta$. Within each of these three groups, we select the worst $\bm{t}_{worst}$ having the minimal $\hat{F}$ among the designs of the same efficiency level (e.g., those with $\text{eff}_{\hat{F}} \geq 0.95$). Note again that different pilot-study designs will give different $\hat{F}$, resulting in different $\bm{t}_{worst}$. For each $\bm{t}_{worst}$, we evaluate its true $F$-efficiency (i.e. $\text{eff}_{F}(\bm{t})=F(\bm{t})/F(\bm{t}^*)$ with $\bm{t}^*$ optimising the true $F$) under the same $10\times20$ simulation settings described above. This allows us  to  compare the performance of different pilot-study designs in facilitating the search of the optimal design $\bm{t}_{opt}$, when some general search algorithm is employed. We also repeat the same procedure by replacing the worst-case design $\bm{t}_{worst}$ with the median-case design $\bm{t}_{median}$. Specifically, the design $\bm{t}_{median}$ has the median $\hat{F}$ among the group of designs of the same efficiency level.

Figures \ref{fig:Exh_99_worst} shows the grouped box plots of $\text{eff}_F(\bm{t}_{worst})$ with $\theta=0.99$ The box plots for the other $\theta$-values can be found in the supplementary document. Similar to Figure \ref{fig:PSS result}, each box plot is formed by the 20 averaged $\text{eff}_{F} (\bm{t}_{worst})$ from the 20 designs for each design type, and the average is taken over the 10 functional data sets. 

It can be observed that across different values of $\theta$, our hybrid design consistently outperforms the other two designs across different numbers of subjects in terms of both median and interquartile range of the worst-case efficiency $\text{eff}_F(\bm{t}_{worst})$. It is notable that there is a decreasing trend in the efficiency as the number of subjects decreases for all of the three designs. Moreover, as the efficiency threshold $\theta$ decreases, the range of the worst-case efficiencies shifts downward. Nonetheless, our hybrid design still outperforms the other two designs.

A similar result to that of $\bm{t}_{worst}$ is observed for $\bm{t}_{median}$. This is shown in Figure \ref{fig:Exh_99_med} for $\theta=0.99$, and in the supplementary document for the other $\theta$'s. Our hybrid design also consistently surpasses the other two designs in terms of both the median and interquartile range of the median efficiency $\text{eff}_F(\bm{t}_{median})$. 

\begin{figure}[h!]
\centering
\setcounter{figure}{2}
\includegraphics[width=.90\textwidth]{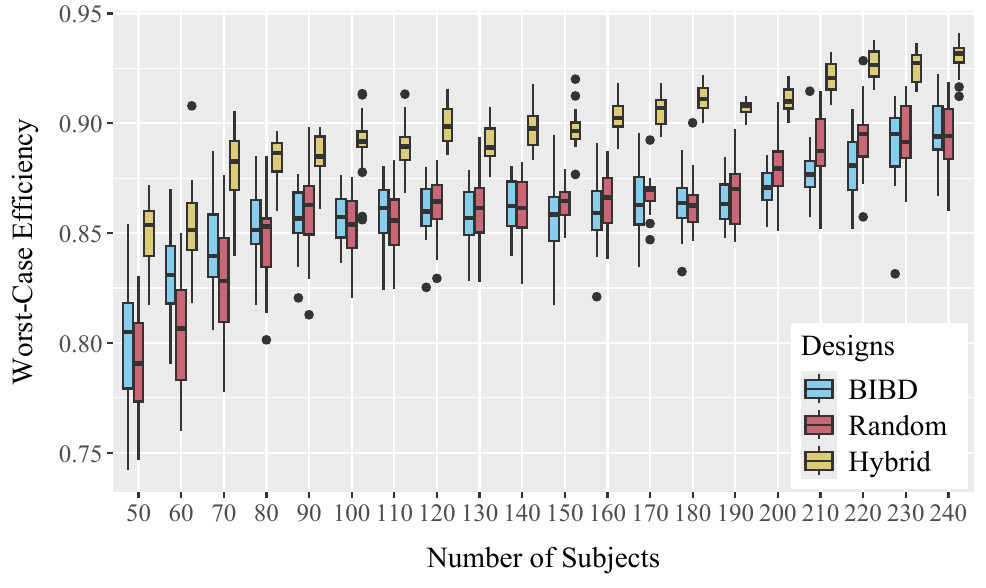}
  \caption{Grouped box plot of designs with $\text{eff}_{F} (\bm{t}_{worst})$ and $\theta = 99\%$.}
  \label{fig:Exh_99_worst}
\end{figure}

\begin{figure}[h!]
\centering
\includegraphics[width=.90\textwidth]{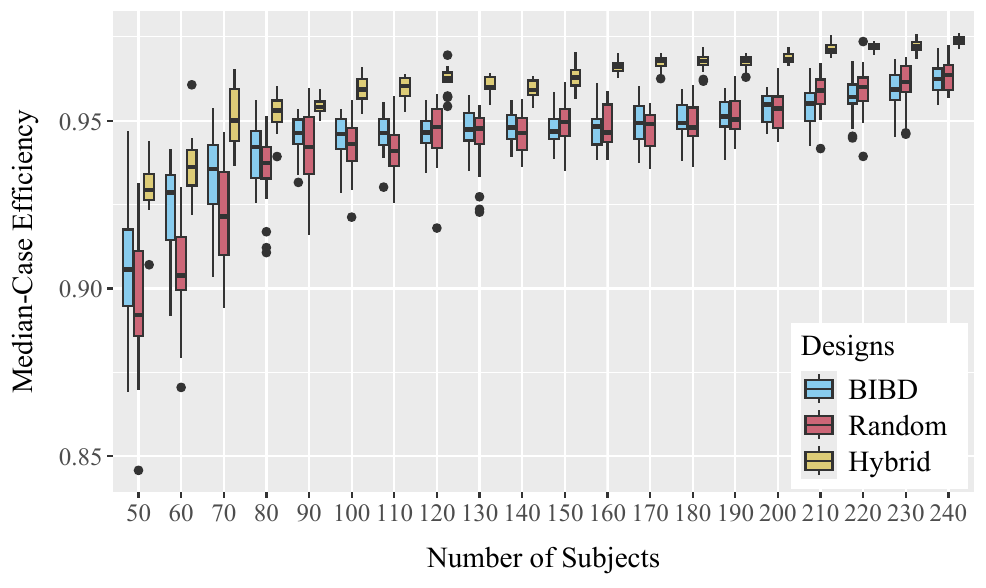}
  \caption{Grouped box plot of designs with $\text{eff}_{F} (\bm{t}_{median})$ and $\theta = 99\%$.}
  \label{fig:Exh_99_med}
\end{figure}

Case Study: Age-Related Patterns in Reproduction \label{sec real data}
In addition to the simulation, we study the performance of our hybrid design on a real data set. The data of interest is one of the most frequently used longitudinal data set in FDA studies collected by \cite{carey1998relationship} and is publicly available through the \texttt{fdapace} package in R. It records the daily number of eggs laid by each of 789 female fruit flies, observed from the day 1 to day 25.  While the data is densely observed, its collection involved substantial labour of daily manual observations of fruit fly eggs, making it a representative case of resource-intensive data acquisition. Such settings highlight the practical need for efficient pilot-study designs that can guide sparse data collection without sacrificing inferential accuracy. 

Similar to Section \ref{sec simulation}, we will compare the performances of three pilot-study design structures in terms of facilitating the identification of high-quality FDA designs and recovery of underlying trajectories for subjects in pilot study. Yet, different from our previous simulation settings, the true parameters of the underlying trajectories in this data set are unknown and may not follow the Gaussian assumption, which invites potential uncertainty from the real world to test our design robustness.

In terms of constructing FDA data sets, we set the subject size as $n = 50, \ldots, 240$. For each $n$, we sample subjects without replacements from the original fruit fly data set to obtain 10 functional data sets. To test the stability of designs, we generate 20 designs for each design structure and sparsify the data sets accordingly. Therefore, in total, for each functional data set and each design structure, we have 20 sparse pilot data sets. 

After obtaining the sparsified pilot data sets, we apply the PACE method introduced in Section \ref{sec model and design} and obtain parameter estimates. With the estimates, we calculate the composite criterion in \eqref{eq:composite criterion}. It is worth noting that the true underlying parameters are unknown, and hence, so is $F(\cdot)$. We thus estimate $F(\cdot)$ from the dense data set of size $789$, and use this estimate as the true $F(\cdot)$ in \eqref{eq:ARE}.  Then, we further evaluate the efficiency of different pilot-study designs under the situation where an arbitrary search algorithm is used to obtain the design for the next study. We assume that the search algorithm at least attains a certain efficiency thresholds $\theta \in \{99\%, 97\%, 95\%\}$. The design efficiencies are computed based on $\text{eff}_{F} (\bm{t})= F(\bm{t})/ F(\bm{t}^*)$, where $\bm{t}^*$ now represents the optimal design obtained with the true dense data set. The efficiencies are evaluated at the worst-case design $\bm{t}_{worst}$.

\begin{figure}[h!]
\centering
\includegraphics[width=.8\textwidth]{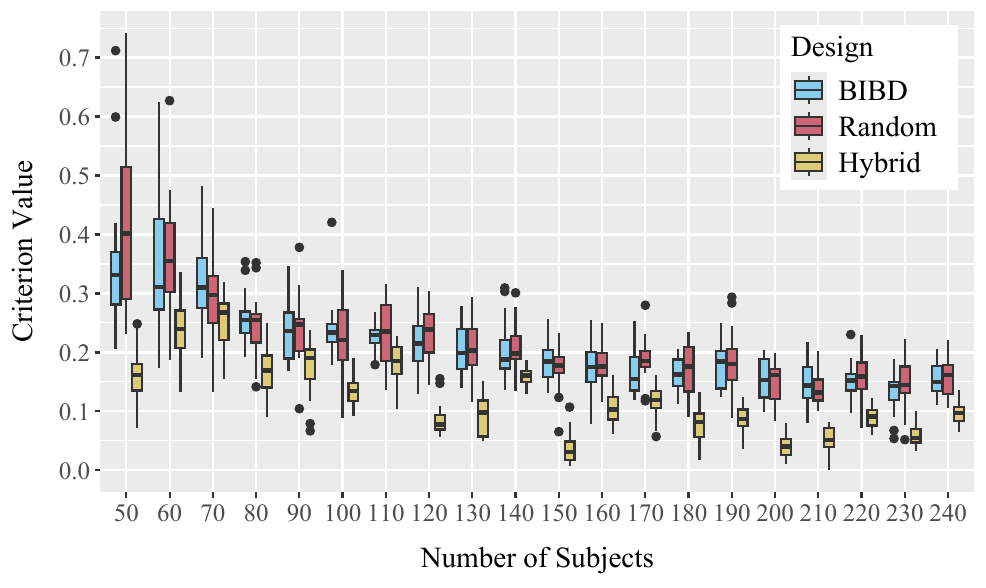}
  \caption{Grouped box plot of designs for composite criterion with real data set.}
  \label{fig:Med_PSS}
\end{figure}

For each number of subjects that we consider, we present in Figure \ref{fig:Med_PSS} the box plots of the criterion values for the three design structures. Each box summarises the averaged criterion values over the 10 functional data sets for the 20 designs for each design structure. For all the scenarios we studied in Figure \ref{fig:Med_PSS} , the performances of our hybrid design consistently surpasses the other two designs in terms of both the median and interquartile range. While the BIBD and random design have similar median values, random design normally has a larger interquartile range. It is noteworthy that the median and interquartile ranges of all designs tend to increase as the number of subjects decreases. 

\begin{figure}[h!]
\centering
\includegraphics[width=.8\textwidth]{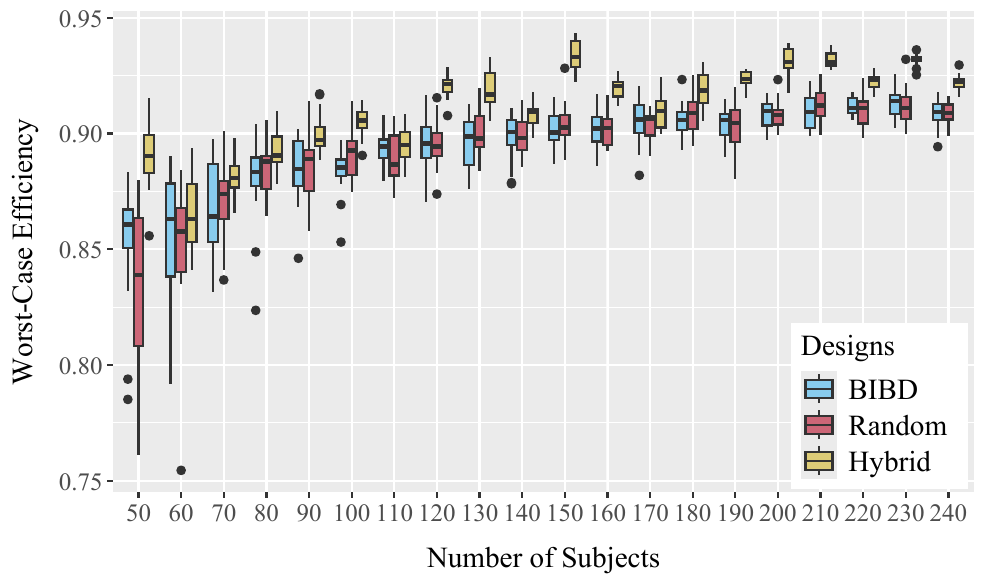}
  \caption{Grouped box plot of designs with $\text{eff}_{F} (\bm{t}_{worst})$ and $\theta = 99\%$ with real data set.}
  \label{fig:Med_worst_99}
\end{figure}

Figures \ref{fig:Med_worst_99} shows the grouped box plots of $\text{eff}_F(\bm{t}_{worst})$ with $\theta=0.99$ The box plots for the other $\theta$-values can be found in the supplementary document. Similar to Figure \ref{fig:Med_PSS}, each box plot is formed by the 20 averaged $\text{eff}_{F} (\bm{t}_{worst})$ from the 20 designs for each design type, and the average is taken over the 10 functional data sets. It can be observed that across different values of $\theta$, our hybrid design consistently outperforms the other two designs across different numbers of subjects in terms of both median and interquartile range of the worst-case efficiency $\text{eff}_F(\bm{t}_{worst})$. It is notable that there is a decreasing trend in the efficiency as the number of subjects decreases for all of the three designs. Moreover, as the efficiency threshold $\theta$ decreases, the range of the worst-case efficiencies shifts downward. Nonetheless, our hybrid design still outperforms the other two designs. Our hybrid design tends to maintain a relatively stable design efficiency when the variation of the criterion values for BIBD and random designs becomes larger as subject size decreases.

\section{Discussion} \label{sec:discussion}
In light of the limitation posed by sparse functional data, we propose a pilot-study design that combines advantages of snippet designs and BIBDs. The snippet design concentrates on the diagonal band of the design plot, and tends to facilitate the estimation of error variance. On the other hand, the BIBD gives a uniformly space-filling design plot with no missing information and enhances the estimation of covariance function. Combining the best of two worlds, our design gives the best time points for collecting data from subjects in the pilot study. It not only facilitates the search of a good design for subjects in the next study but also yields low mean integrated squared errors in recovering random functions in the pilot study. We develop a search algorithm to generate our hybrid design based on linear integer programming. By applying to a real data set, we show that our hybrid design outperforms other extant designs under different scenarios and is relatively stable even when the number of subjects is relatively small.

Beyond its empirical performance, this work highlights the practical importance of incorporating a pilot design stage in functional data studies. In settings where data collection is labour intensive or financially costly, even a modest investment in design can lead to substantial improvements in efficiency and inferential quality. The applications introduced earlier, including biological and clinical studies as well as environmental and high-cost measurement settings, underscore the broad relevance of this approach. By promoting more informed and resource-conscious planning, our design supports the development of scalable and sustainable strategies for functional data collection.

\bibliographystyle{apalike}
\bibliography{reference}

\begin{thebibliography}{}

\bibitem[Carey et~al., 1998]{carey1998relationship}
Carey, J.~R., Liedo, P., M{\"u}ller, H.-G., Wang, J.-L., and Chiou, J.-M.
  (1998).
\newblock Relationship of age patterns of fecundity to mortality, longevity,
  and lifetime reproduction in a large cohort of mediterranean fruit fly
  females.
\newblock {\em The Journals of Gerontology Series A: Biological Sciences and
  Medical Sciences}, 53(4):B245--B251.

\bibitem[Galbraith et~al., 2017]{galbraith2017accelerated}
Galbraith, S., Bowden, J., and Mander, A. (2017).
\newblock Accelerated longitudinal designs: An overview of modelling, power,
  costs and handling missing data.
\newblock {\em Statistical methods in medical research}, 26(1):374--398.

\bibitem[Gregory et~al., 2024]{gregory2024scalable}
Gregory, W., MacEachern, R., Takao, S., Lawrence, I.~R., Nab, C., Deisenroth,
  M.~P., and Tsamados, M. (2024).
\newblock Scalable interpolation of satellite altimetry data with probabilistic
  machine learning.
\newblock {\em Nature Communications}, 15(1):7453.

\bibitem[Ji and M{\"u}ller, 2017]{ji2017optimal}
Ji, H. and M{\"u}ller, H.-G. (2017).
\newblock Optimal designs for longitudinal and functional data.
\newblock {\em Journal of the Royal Statistical Society Series B: Statistical
  Methodology}, 79(3):859--876.

\bibitem[Kodikara et~al., 2022]{kodikara2022statistical}
Kodikara, S., Ellul, S., and L{\^e}~Cao, K.-A. (2022).
\newblock Statistical challenges in longitudinal microbiome data analysis.
\newblock {\em Briefings in Bioinformatics}, 23(4):bbac273.

\bibitem[Lin et~al., 2021]{lin2021basis}
Lin, Z., Wang, J.-L., and Zhong, Q. (2021).
\newblock Basis expansions for functional snippets.
\newblock {\em Biometrika}, 108(3):709--726.

\bibitem[Lopes et~al., 2021]{lopes2021real}
Lopes, M.~B., Tu, C., Zee, J., Guedes, M., Pisoni, R.~L., Robinson, B.~M.,
  Foote, B., Hedman, K., James, G., Lopes, A.~A., et~al. (2021).
\newblock A real-world longitudinal study of anemia management in
  non-dialysis-dependent chronic kidney disease patients: a multinational
  analysis of ckdopps.
\newblock {\em Scientific reports}, 11(1):1784.

\bibitem[Mandal et~al., 2014]{mandal2014efficient}
Mandal, B., Gupta, V., and Parsad, R. (2014).
\newblock Efficient incomplete block designs through linear integer
  programming.
\newblock {\em American Journal of Mathematical and Management Sciences},
  33(2):110--124.

\bibitem[Pan et~al., 2023]{pan2023reinforced}
Pan, Y., Laber, E.~B., Smith, M.~A., and Zhao, Y.-Q. (2023).
\newblock Reinforced risk prediction with budget constraint using irregularly
  measured data from electronic health records.
\newblock {\em Journal of the American Statistical Association},
  118(542):1090--1101.

\bibitem[Park et~al., 2018]{park2018joint}
Park, S.~Y., Xiao, L., Willbur, J.~D., Staicu, A.-M., and Jumbe, N. (2018).
\newblock A joint design for functional data with application to scheduling
  ultrasound scans.
\newblock {\em Computational Statistics \& Data Analysis}, 122:101--114.

\bibitem[Ramsay, 1982]{ramsay1982data}
Ramsay, J.~O. (1982).
\newblock When the data are functions.
\newblock {\em Psychometrika}, 47:379--396.

\bibitem[Rha et~al., 2020]{rha2020design}
Rha, H., Kao, M.-H., and Pan, R. (2020).
\newblock Design optimal sampling plans for functional regression models.
\newblock {\em Computational Statistics \& Data Analysis}, 146:106925.

\bibitem[Rice and Silverman, 1991]{rice1991estimating}
Rice, J.~A. and Silverman, B.~W. (1991).
\newblock Estimating the mean and covariance structure nonparametrically when
  the data are curves.
\newblock {\em Journal of the Royal Statistical Society: Series B
  (Methodological)}, 53(1):233--243.

\bibitem[Shi et~al., 2021]{shi2021functional}
Shi, H., Dong, J., Wang, L., and Cao, J. (2021).
\newblock Functional principal component analysis for longitudinal data with
  informative dropout.
\newblock {\em Statistics in Medicine}, 40(3):712--724.

\bibitem[Silverman, 1985]{silverman1985some}
Silverman, B.~W. (1985).
\newblock Some aspects of the spline smoothing approach to non-parametric
  regression curve fitting.
\newblock {\em Journal of the Royal Statistical Society: Series B
  (Methodological)}, 47(1):1--21.

\bibitem[Yao et~al., 2005]{yao2005functional}
Yao, F., M{\"u}ller, H.-G., and Wang, J.-L. (2005).
\newblock Functional data analysis for sparse longitudinal data.
\newblock {\em Journal of the American statistical association},
  100(470):577--590.

\bibitem[Yates, 1936]{yates1936incomplete}
Yates, F. (1936).
\newblock Incomplete randomized blocks.
\newblock {\em Annals of eugenics}, 7(2):121--140.

\bibitem[Zhong et~al., 2022]{zhong2022robust}
Zhong, R., Liu, S., Li, H., and Zhang, J. (2022).
\newblock Robust functional principal component analysis for non-gaussian
  longitudinal data.
\newblock {\em Journal of Multivariate Analysis}, 189:104864.

\bibitem[Zhu et~al., 2022]{zhu2022spatiotemporal}
Zhu, W., Zhu, Z., and Dai, X. (2022).
\newblock Spatiotemporal satellite data imputation using sparse functional data
  analysis.
\newblock {\em The Annals of Applied Statistics}, 16(4):2291--2313.

\end{thebibliography}

\end{document}